# Microwave and Acoustic Absorption Metamaterials

Sichao Qu[*], Ping Sheng[†]

*Department of Physics, The Hong Kong University of Science and Technology*

*Clear Water Bay, Kowloon, Hong Kong, China*

Wave absorption metamaterials have been an enduring topic over the past two decades, propelled not only by novel scientific advances, but also by their extensive application potential. In this review, we aim to provide some general insights into the absorption mechanism common to both microwave and acoustic systems. By establishing a universal model for resonance-based metamaterials, we present the theoretical conditions for broadband impedance matching and introduce the fundamental causal limit as an evaluation tool for absorption performance. Under this integrated framework, we survey the recent advances on metamaterials absorption in both microwave and acoustic systems, with the focus on those that pushed the overall performance close to the causal limit. We take note of some new, emerging metastructures that can circumvent the constraint imposed by causal limit, thereby opening a new avenue to low-frequency absorption. This review concludes by discussing the existing challenges with possible solutions and the broad horizon for future developments.

## I. INTRODUCTION

Wave absorption is important to many aspects of our daily life. We rely on ultraviolet absorption by atmospheric molecules to protect us from skin damage. The solar cells absorb the sunlight through the photovoltaic mechanism to provide green energy. Capture and processing of the ubiquitous microwave signals are now the basis of high-speed communication that is indispensable in our society. We install sound-absorbing materials in buildings to alleviate the noise problems, and/or improve the indoor auditory experience. In view of such diverse significance of wave absorption, the development of high-performance broadband absorbers has long been a widely pursued topic. However, for traditional materials [1, 2], such as porous absorptive foam, there exist inherent restrictions on material properties, thereby leading to the low tunability of the absorption spectrum. Thanks to the revolutionary development of the metamaterials [3-9], which are artificial composite structures that rely on the interaction between the structural geometry, material characteristics, and wave properties for the resulting wave manipulation capabilities, we are now presented with myriad possibilities to tune the metamaterial functionalities that cannot be found in nature. In particular, it becomes possible to customize the absorption spectrum according to the specific targets in different application scenarios.

An essential element in the absorber design is impedance matching in the desired frequency band to avoid reflection from the sample. However, it turns out that associated with impedance matching there is another important consideration—sample thickness. In practice, it is always a challenge to attain high dissipation in the low-frequency range within a given wavelength, because for any linear response system, the dissipation coefficient is a quadratic function of frequency [10]. Even if some metamaterials can take advantage of the low-frequency modes to achieve excellent absorption with a thin sample thickness, we will see that their operating bandwidth is usually extremely narrow. Therefore, a natural question arises: is there some general constraint between the absorption spectrum and the sample thickness? The answer turns out to be "yes" for arbitrary linear, passive, and time-invariant systems. Theoretical studies revealed that the integral of the reflection coefficient in the logarithmic dB scale, concerning wavelength, represents a minimum sample thickness (apart from some constants) that must be smaller than the actual sample thickness. This well-known relation is called *causal limit* or *causality constraint* [11, 12]. Therefore, for a given available space, we can know in advance the maximal absorption performance over the desired frequency band [13]. One can also turn the causal limit into an effective evaluation tool [14] for optimizing absorber performance to approach the *causal optimality* through iterations of the absorber design parameters.

The integrated consideration of impedance matching and causal limit provides an objective perspective for this review article to survey the microwave and acoustic absorption metamaterials. In what follows, we first make concrete the theoretical framework in Section II by superposing Lorentzian functions, which can be used to analyze any linear wave systems with the causal response. Based on this framework, we derive the theoretical conditions for an absorber to achieve broadband impedance matching condition while simultaneously also approaching the causal limit. This is followed by a survey of the representative works in the microwave and airborne acoustic absorption systems in Section III. In particular, we focus on the transition from local resonances to broadband absorption along the development trajectory of metamaterial absorbers. Defined by the causal limit, optimal broadband absorbers are observed to emerge with diverse design approaches. Section IV notes the very recent works that can mitigate/circumvent the restriction imposed by the original causal limit. We conclude in Section V by delineating the challenges faced by metamaterial absorbers in the transition to products and the broad horizons when such challenges can be met.

[*] Corresponding email 1: squ@connect.ust.hk;

[†] Corresponding email 2: sheng@ust.hk;

## II. A THEORETICAL MODEL FOR RESONANCE-BASED METAMATERIALS

### A. Model definitions

Due to the subwavelength feature of the metamaterials, we model wave absorption phenomenon as a one-dimensional problem (i.e., normal incidence) for simplicity. This is justified because even at oblique incidence, the phase difference across the subwavelength sample surface is small, as long as the incidence angle is not close to 90 degrees. To unify the description of both the microwave and acoustic systems, we adopt the following mapping [15]: $\phi \rightarrow H, p$ for magnetic field and acoustic pressure modulation, respectively; while the counterpart to $\phi$ is $\varphi$, which denotes either the electric field $E$ or acoustic (particle) displacement velocity $v$. It is understood that the electromagnetic (EM) fields are transverse (TEM mode), and the acoustic field $p$ is scalar in character, with a longitudinal velocity field $v$. We note that both $\phi_i$ and $\phi_r$, where the subscripts $i$, $r$ denote incidence and reflection, respectively, are plane waves with the phase variation $e^{i(kx-\omega t)}$, where $k, \omega$ denote the wavevector and angular frequency, respectively, while $x$ represents the direction of wave propagation, and $t$ is time.

In Fig.1(a), consider an incident field $\phi_i$ incident on the interface between the air and metamaterials, defined to be at $x = 0$. Imperfect impedance matching would cause a backward reflected field $\phi_r$. To eliminate transmission, we specify a Neumann boundary condition $\partial_x \phi|_{x=d} = 0$ on the backside of the sample (i.e., perfect electric conductor boundary for EM case and hard boundary for the acoustic case). The reflection coefficient is given by

$$R = \frac{\phi_r}{\phi_i}\bigg|_{x=0} = \frac{Z_s - Z_c}{Z_s + Z_c}, \tag{1}$$

where $Z_c$ denotes the generalized impedance of the incident medium, i.e., the characteristic admittance $\sqrt{\epsilon_0/\mu_0}$ in the EM case, or characteristic impedance $\sqrt{\rho_0 B_0}$ in the acoustic case. Here $\epsilon_0$, $\mu_0$ denote permittivity and permeability in EM case while $\rho_0$, $B_0$ denote mass density and bulk modulus in acoustic case. And the generalized surface impedance $Z_s$ is defined by

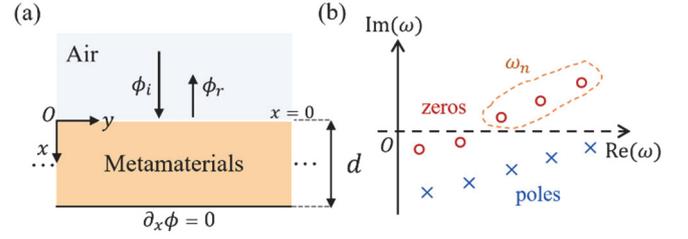

**Fig. 1 (a)** Schematic illustration of the metamaterial absorbing layer, modelled as a one-dimensional problem. $\phi_i$ is the incident field and $\phi_r$ is the reflected field. The total field $\phi = \phi_i + \phi_r$. **(b)** The complex frequency plane of $\ln(|R|)$ with a schematic illustration for the locations of its zeros (red circles) and poles (blue crosses). Dots encircled by the dashed line are the set of $\omega_n$ defined on the upper complex plane, where $\text{Im}(\omega_n) > 0$.

$$Z_s = \frac{\phi}{\varphi}\bigg|_{x=0}, \tag{2}$$

where both $\phi$ and $\varphi$ are the total fields (summation of forward and backward waves). In the absence of transmission, the absorption coefficient is given by $A = 1 - |R|^2$ according to energy conservation, which can be further expressed as

$$A = 1 - \left|\frac{Z_s - Z_c}{Z_s + Z_c}\right|^2. \tag{3}$$

Impedance matching is realized when $Z_s = Z_0$, at which $A = 1$. For the ease of ensuing theory formulation, we introduce the Green function $G(x, x')$ [15, 16] to characterize the response of metamaterials. In particular, by investigating the surface response ($x = x' = 0$), we can link $Z_s$ with $G(\omega)$ by the formula:

$$G(\omega) = \frac{1}{-i\omega Z_s}. \tag{4}$$

From Eq. (4), we learn that the surface impedance $Z_s$ can be obtained once $G(\omega)$ is known.

### B. Green function for multiple resonances

Within a metamaterial, multiple resonant states can be excited by an incident wave. By projecting these states onto the material-air interface, we can obtain the overall response, given by the superposition of the Lorentzian functions

$$G(\omega) = \sum_{i=1}^{N} \frac{\alpha_i}{\omega_i^2 - \omega^2 - i\beta_i \omega}, \tag{5}$$

where $\omega_i$, $\alpha_i$ and $\beta_i$ are the $i^{\text{th}}$ resonance frequency, oscillation strength and dissipation factor, respectively. The coefficients ($\omega_i$, $\alpha_i$ and $\beta_i$) in $G(\omega)$ can be expressed in terms of the eigenfunctions of the absorption sample, which may be evaluated either analytically when possible, or numerically when analytical solutions are difficult to obtain. The eigenfunctions are defined to be those obtained by setting the Dirichlet boundary condition on the sample surface thus being non-radiating in nature. For related examples of mode expansion theory, one can refer to Ref. [17] and Ref. [15] for EM and acoustic metamaterials, respectively.

We note that since each Lorentzian function follows the Kramers-Kronig relations [18], which is derived based on the assumption of linearity and causality, the summation $G(\omega)$ must also inherit these properties. In fact, Dirdal et al. [19] rigorously proved that all causal functions can be approximated by the superpositions of Lorentzian functions to arbitrary precision. Therefore, starting from Eq. (5), our model is universal and can be used to describe any causal wave system, which serves as one of the theoretical bases of our following analysis of the resonance-based metamaterials.

## C. Causal limit as an evaluation tool

For any linear and passive absorber, its reflected wave $\phi_r(t)$ can only depend on the incident wave before the current time instant $t$. Mathematically, it can be formulated in the form of a convolution integral:

$$\phi_r(t) = \int_0^\infty K(\tau)\phi_i(t-\tau)d\tau, \tag{6}$$

where $K(\tau)$ is the response kernel function in the time domain, whose Fourier transform gives the reflection coefficient in the frequency domain, i.e., $R(\omega) = \mathcal{F}\{K(\tau)\}$. By analyzing the analyticity of a derivative function containing $R(\omega)$, we can obtain the following relation [11, 12, 20]:

$$-\int_0^\infty \ln|R(\omega)|\frac{d\omega}{\pi\omega^2} = \frac{d}{c_0 F} - \sum_n \frac{\text{Im}(\omega_n)}{|\omega_n|^2}. \tag{7}$$

Here $d$ is the thickness of the absorber, $c_0$ is the wave speed in the ambient environment, and $\omega_n$ denotes the zeros of $\ln(|R|)$ that are located in the upper half-plane of complex frequency plane, i.e., $\text{Im}(\omega_n) > 0$ [see Fig.1(b)]. For EM and acoustic cases, the dimensionless factor $F$ is defined by

$$F = \begin{cases} \dfrac{B_{\text{eff}}}{B_0} & \text{(Acoustics)} \\[2ex] \dfrac{\mu_0}{\mu_{\text{eff}}} & \text{(EM)} \end{cases}, \tag{8}$$

with $\mu_{\text{eff}}$, $B_{\text{eff}}$ being the effective permeability and effective bulk modulus of the metamaterials at the static limit $\omega \to 0$. Since the summation term in Eq.(7) is always positive, we can alternatively turn Eq.(7) into an inequality, known as the *causal limit*

$$d \geq d_{\min} = \frac{F}{4\pi^2}\left|\int_0^\infty \ln|1 - A(\lambda)|\, d\lambda\right|, \tag{9}$$

where we have substituted reflection $R$ by absorption $A$ and angular frequency $\omega$ by the wavelength $\lambda$. $d_{\min}$ is noted to be the minimal thickness dictated by the causal limit. We learn from Eq. (9) that perfect absorption for all frequencies is impossible with a finite sample thickness, i.e., the integral does not converge. Also, to some extent, Eq. (9) explains why the low-frequency absorption is difficult with a thin sample because the long-wavelength part of the integral is dominant in determining the value of $d_{\min}$.

We introduce a causality ratio [14] as a tool for evaluating the absorption performance, given by

$$R_c = \frac{d}{d_{\min}}. \tag{10}$$

If $R_c = 1$, all the zeros [red circles in Fig. 1(b)] must shift to the lower plane and thus the absorber is defined to be causally '*optimal*' [14, 20]. We will show this condition to be achievable by increasing either the dissipation or the mode density. It should be noted that if there exists a zero exactly on the real frequency axis in Fig. 1(b), such a condition is regarded as *critical coupling* [21], which, however, is not the necessary condition for causal optimality.

We emphasize that when evaluating the performance of an absorber, the evaluation index $R_c$ should be used together with the operating bandwidth, since single band absorber can also realize $R_c = 1$ (see examples in Ref. [13]). Therefore, in what follows, we will derive the condition for broadband impedance matching while exploring how to approach the causal optimality simultaneously.

## D. Theoretical conditions for broadband impedance matching

Consider an idealized collection of resonances with a continuum distribution of resonance frequencies, i.e., $\Omega \in [\omega_1, \infty)$. We define the mode density to be $D(\Omega) = dN/d\omega$, where $dN$ is the mode number within the frequency interval $[\omega, \omega + d\omega]$. Then Eq. (5) can be alternatively formulated as

$$G(\omega) = \int_{\omega_1}^\infty \frac{\alpha(\Omega)D(\Omega)}{\Omega^2 - \omega^2 - i\omega\beta(\Omega)}d\Omega, \tag{11}$$

which can be divided into real and imaginary parts as follows. Based on the small dissipation assumption (i.e., $\beta \ll \omega$), we can utilize the asymptotic definition of the Dirac delta function [22] to obtain

$$\begin{aligned} G(\omega) = P\int_{\omega_1}^\infty \frac{\alpha(\Omega)D(\Omega)}{\Omega^2 - \omega^2}d\Omega \\ + i\pi\int_{\omega_1}^\infty \alpha(\Omega)D(\Omega)\delta(\Omega^2 - \omega^2)d\Omega \end{aligned} \tag{12}$$

The principal value of the first term on the right-hand side, can be approximately treated as zero for $\omega > \omega_1$. Our present target is to attain impedance matching over a broadband frequency range above $\omega_1$. According to Eq. (3), this requires $\mathrm{Im}[G(\omega)] \cong 1/(\omega Z_c)$. Therefore, by combining this target with Eq. (12), we have the *impedance matching condition* for the continuum model

$$\alpha(\omega)D(\omega) = \frac{2}{\pi Z_c}. \tag{13}$$

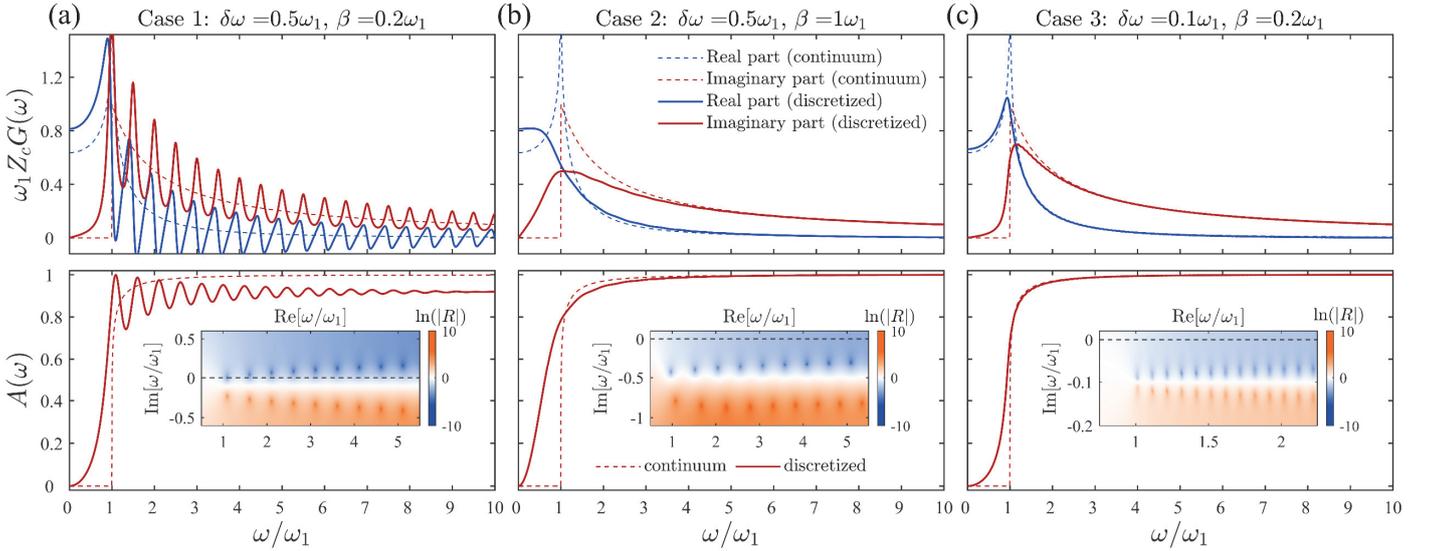

**Fig. 2** The Green function (in dimensionless form given by $\omega_1 Z_c G(\omega)$) and the corresponding absorption spectrum $A(\omega)$. In the panels, the dashed lines denote the results from continuum resonance model, given by Eq. (15), that serves as the causal target. The solid lines correspond to the discretized resonance model, given by Eq. (5). The insets are the value of $\ln(|R|)$ on the complex frequency plane. The blue and orange dots are the zeros and poles, respectively. **(a)** Case 1: the reference case with relatively low mode density and low dissipation factor ($\delta\omega = 0.5\omega_1, \beta = 0.2\omega_1$); **(b)** Case 2: low mode density and high dissipation factor ($\delta\omega = 0.5\omega_1, \beta = \omega_1$); **(c)** Case 3: high mode density and low dissipation factor ($\delta\omega = 0.1\omega_1, \beta = 0.2\omega_1$).

By substituting Eq.(13) back into Eq.(11), an analytical form for idealized continuous resonances can be derived from

$$G(\omega) = \lim_{\beta \to 0} \int_{\omega_1}^{\infty} \frac{2/(\pi Z_c)}{\Omega^2 - \omega^2 - i\beta\omega} d\Omega, \tag{14}$$

which can be further reduced to

$$G(\omega) = \begin{cases} \frac{1}{\omega\pi Z_c}\ln\left(\frac{\omega_1 + \omega}{\omega_1 - \omega}\right), & \text{if } \omega < \omega_1 \\ \frac{1}{\omega Z_c}\left[i + \frac{2}{\pi}\tanh^{-1}\left(\frac{\omega_1}{\omega}\right)\right], & \text{if } \omega > \omega_1 \end{cases} \tag{15}$$

which is plotted as dashed lines in the upper panels of Fig.2. In Eq. (15), the disappearance of $\mathrm{Im}(G)$ below $\omega_1$ is due to the assumed non-existence of mode density below $\omega_1$, thus the absorption must also be zero.

In actual practice, we can only have a finite number of discretized resonances to approach the causal target given by Eq. (15). Hence in this case we define the mode density to be $D(\omega_i) = 1/(\omega_{i+1} - \omega_i) = 1/\delta\omega_i$ and the *impedance matching condition* in Eq. (13), can be modified to fit the discretized model, which writes

$$\alpha_i = \frac{2\delta\omega_i}{\pi Z_c}. \tag{16}$$

From Eq. (13) and Eq. (16), we learn that the product of the oscillation strength and mode density should be a constant in order to best attain the broadband impedance matching. While the above condition is derived for the target of broadband impedance matching above a cutoff frequency $\omega_1$, the present approach can be extended to construct the conditions for attaining arbitrary broadband impedance [23, 24].

The above shows theoretically that if an absorber, with a given sample thickness and absorption spectrum, deviates from the $R_c = 1$ condition, then there is potential for improving the absorption without increasing the sample thickness. Usually that is difficult for the traditional absorbers since it may involve changing the material properties. However, this is precisely the area where metamaterials can help, by tuning the mode density of the resonances and the dissipation coefficient associated with each resonance.

In what follows, we demonstrate, by using the model of broadband total absorption above a cutoff frequency, how the two parameters—resonance mode density and dissipation coefficient—can be utilized to *approach* the $R_c = 1$ condition. Subsequently, we show through a survey of the metamaterials in literature how the development of the metamaterial absorbers—from narrow frequency to broadband—has roughly followed the above framework.

### E. Relation between the thickness and the static Green function

In order to consider the model of discretized resonances, we first show, as the reference, the consistency of the continuum model used in the previous sections. To this end, we build up the relation between the absorber's thickness $d$ and the static Green function, for the purpose of calculating $R_c$. We have assumed a Neumann boundary condition $\partial_x \phi|_{x=d} = 0$ on the backside of the sample. Therefore, according to the impedance transfer formula, the surface impedance of the absorber is

$$Z_s = i Z_{\text{eff}} \cot\left(\omega d / c_{\text{eff}}\right), \tag{17}$$

where $Z_{\text{eff}}$, $c_{\text{eff}}$ denote the generalized effective impedance and effective wave speed of the metamaterials, respectively. By inserting Eq. (17) into Eq. (4), we have

$$G(\omega) = \frac{\tan\left(\omega d / c_{\text{eff}}\right)}{\omega Z_{\text{eff}}}, \tag{18}$$

whose static limit allows us to extract the thickness

$$d = c_{\text{eff}} Z_{\text{eff}} \lim_{\omega \to 0} G(\omega), \tag{19}$$

where $c_{\text{eff}} Z_{\text{eff}} = B_{\text{eff}}$ in acoustic case and $c_{\text{eff}} Z_{\text{eff}} = 1/\mu_{\text{eff}}$ in EM case. Based on Eq. (19), we compare $d$ with $d_{\min}$, and thereby $R_c$ can be obtained.

## F. Evaluation of $R_c$ for continuum and discretized cases

We show analytically that for idealized continuum case with the causal impedance form [see Eq. (15)], causal optimality always holds, i.e., $R_c = 1$. However, for discretized case, which is always the practical case, we numerically demonstrate that the condition $R_c = 1$ is conditional, depending on the locations of the zeros of $\ln(|R|)$ on the complex frequency plane.

*Idealized continuum case.* We revisit the idealized continuum model. By combining Eq. (19) and Eq. (15), we can analytically get the corresponding thickness

$$d = c_{\text{eff}} Z_{\text{eff}} \lim_{\omega \to 0} \frac{1}{\omega \pi Z_c} \ln\left(\frac{\omega_1 + \omega}{\omega_1 - \omega}\right) = \frac{2 c_{\text{eff}} Z_{\text{eff}}}{\pi \omega_1 Z_c} = \frac{2 F c_0}{\pi \omega_1}. \tag{20}$$

The minimal thickness dictated by causal limit can be obtained by inserting Eq. (15) into Eq. (3) and analytically performing the integral in Eq. (9):

$$d_{\min} = \frac{F c_0}{2\pi} \int_{\omega_1}^{\infty} \ln\left|\frac{i \tanh^{-1}\left(\omega_1 / \omega\right)}{\pi - i \tanh^{-1}\left(\omega_1 / \omega\right)}\right| \frac{d\omega}{\omega^2} = \frac{2 F c_0}{\pi \omega_1}, \tag{21}$$

which is exactly the same as that given by Eq. (20). Therefore, the continuum model, based on the impedance matching condition [Eq. (13)], is self-consistent and causally optimal. This can serve as a target, with its corresponding absorption plotted as dashed lines in the lower panels of Fig. 2, that can be approached by the discretized model (solid lines in Fig. 2).

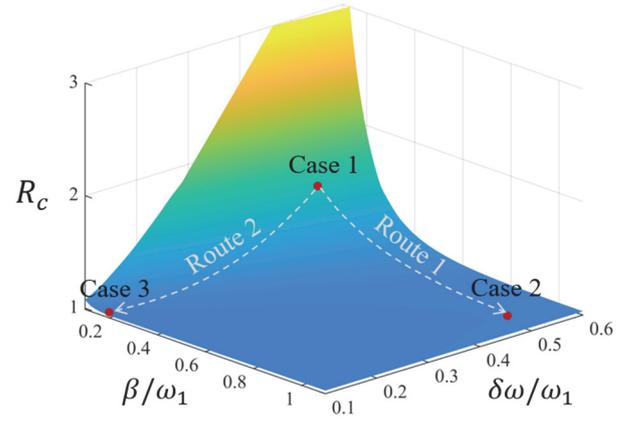

**Fig. 3** The value of the causality ratio, plotted as a function of dimensionless parameters $\beta$ and $\delta\omega_c$. Variation of the two parameters represents two options to achieve causal optimality.

*Discretized case.* We show that causality optimality does not necessarily hold in the discretized case. By inserting Eq. (5) into Eq. (19), we obtain the thickness given by

$$d = \begin{cases} B_{\text{eff}} \sum_{i=1}^{N} \alpha_i / \omega_i^2 & \text{(Acoustics)} \\ \left(1/\mu_{\text{eff}}\right) \sum_{i=1}^{N} \alpha_i / \omega_i^2 & \text{(EM)} \end{cases}. \tag{22}$$

Based on Eq. (22), we compare $d$ with $d_{\min}$, and thereby $R_c$ can be numerically obtained.

For simplicity of demonstration, we assume all the Lorentzian terms in Eq. (5) are with the same dissipation factor ($\beta_i = \beta$) and the resonances are distributed evenly, which means the mode density is constant, given by

$$D(\omega) = \frac{1}{\omega_{i+1} - \omega_i} = \frac{1}{\delta\omega}. \tag{23}$$

Now, we can demonstrate with numerical results since analytical forms are not available. We selected three cases with different ($\delta\omega, \beta$) values (displayed on the top panels of Fig.2), and the oscillation strengths are set accordingly to satisfy the impedance matching condition given by Eq. (16).

For case 1 (the first column), the mode density is relatively low and the dissipation is also small. As a consequence, the Green function deviates from the target with an oscillating manner and the absorption also fluctuates around 0.9. In the inset, it is seen that the zeros of $\ln(|R|)$ are on the upper-frequency plane, indicating the system is not causally optimal. In order to improve the absorption and also to approach the causal limit, we propose two feasible routes. The first one is to increase the dissipation $\beta$ from $0.2\omega_1$ to $\omega_1$ (noted as case 2). From the results displayed in Fig. 2(b), we can see that the Green function is now closer to the target idealized model. Also, the absorption becomes smooth and flat at a high level. The second route is to increase the mode density by changing $\delta\omega$, from $0.5\omega_1$ to $0.1\omega_1$ (case 3). In this way, the Green function and absorption almost coincide with the targets. We can conclude that both routes are feasible for approaching the idealized continuum model and thereby causally optimality [see the insets in Fig. 2(b) and Fig. 2(c)]. In Fig. 3, we scan the $R_c$ values by varying $(\beta, \delta\omega)$ and mark the two routes that we have demonstrated. The blue color indicates the region with a low $R_c$ close to unity.

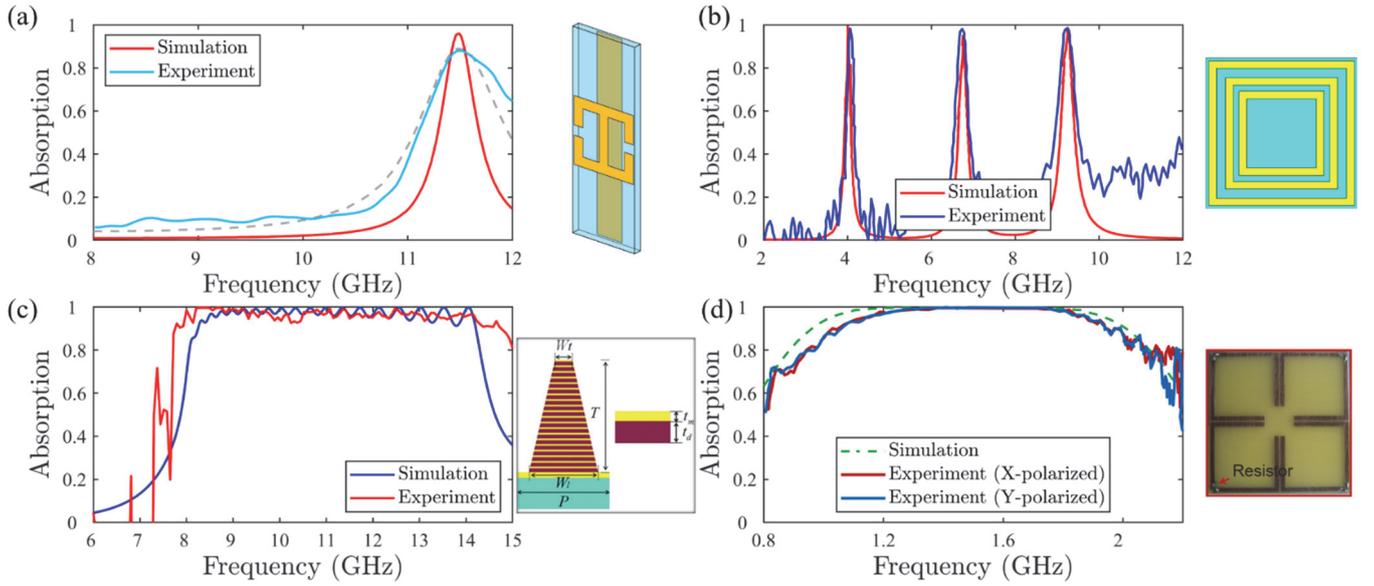

**Fig. 4** The performance and structure of microwave metamaterial absorbers. **(a)** A single-band absorber with high absorption at 11.5 GHz. **(b)** A triple-band absorber with concentric three metal-ring structure. **(c)** A multilayer absorber with a wide absorption band, facilitated by stacked patch resonators. **(d)** A low-frequency wideband absorber equipped with four resistive metallic rings. For **(a)** and **(b)**, sample schematics are plotted from the top view, i.e., the incident wavevector is perpendicular to the plane of the structure. For **(c)** and **(d)**, the samples are viewed from the side, with incident wavevector in the plane of the illustration. Reused with permissions from Ref. [31], Ref. [36], Ref. [40] and Ref. [53].

In summary, the theoretical model gives us several important lessons. First, the impedance matching condition should be satisfied [as Eq. (15)]. In other words, if we regard the modes as harmonic oscillators, the oscillation strength of each one needs to be fitted with the mode density to create a wide absorption frequency range. The second is that the dissipation of the system should also be adapted to the mode density (see the two routes shown in Fig. 3). Although our model is simplified compared to the practical cases, the results are of universal significance for the design of broadband absorbers.

## III. ABSORPTION BY RESONANCE-BASED METAMATERIALS

Early metamaterial absorbers relied on a single or a few resonant modes to concentrate energy density, thus attaining excellent but usually narrowband absorption with deep subwavelength thicknesses, which is not typically feasible from the traditional materials. Meanwhile, the tunability of the absorption frequency is also a unique feature that has attracted strong interest due to its application value [25].

However, with the advances in the understanding of the absorption mechanisms, some structures were designed to break the narrowband nature and extend the absorption ability to a wider frequency range. A transition from narrowband to broader frequency range absorption was an inevitable step in the evolution of the metamaterial absorbers.

### A. Microwave absorbers

Microwave is the most challenging frequency regime for absorption among diverse EM bands, due to its long wavelength and high penetrability through solids. Driven by the need for application related to radars, Dallenbach [26] and Salisbury [27] absorbers were the earliest examples to the best of our knowledge. These early microwave absorbers shared similar absorption mechanisms based on the quarter-wavelength resonances, fitted with a lossy layer. However, such classical absorbers [28-30] are usually thick in low-frequency absorption because of their requirements on quarter wavelength. More recently, the first metamaterial-based absorber was proposed by Landy et al. [31], with its thickness only 1/35 of the relevant wavelength [see Fig. 4(a)]. The split-ring structure was responsible for enhancing the electric couplings while the cut line behind can support magnetic response. In this way, they were able to decouple effective permittivity and permeability and individually tune their respective resonances. However, such narrowband absorption at a single frequency [31-35] may have only limited applications.

In order to extend the absorption frequency bands, one can either increase the resonance mode density (denoted route 1), or the loss of the system (denoted route 2), as previously

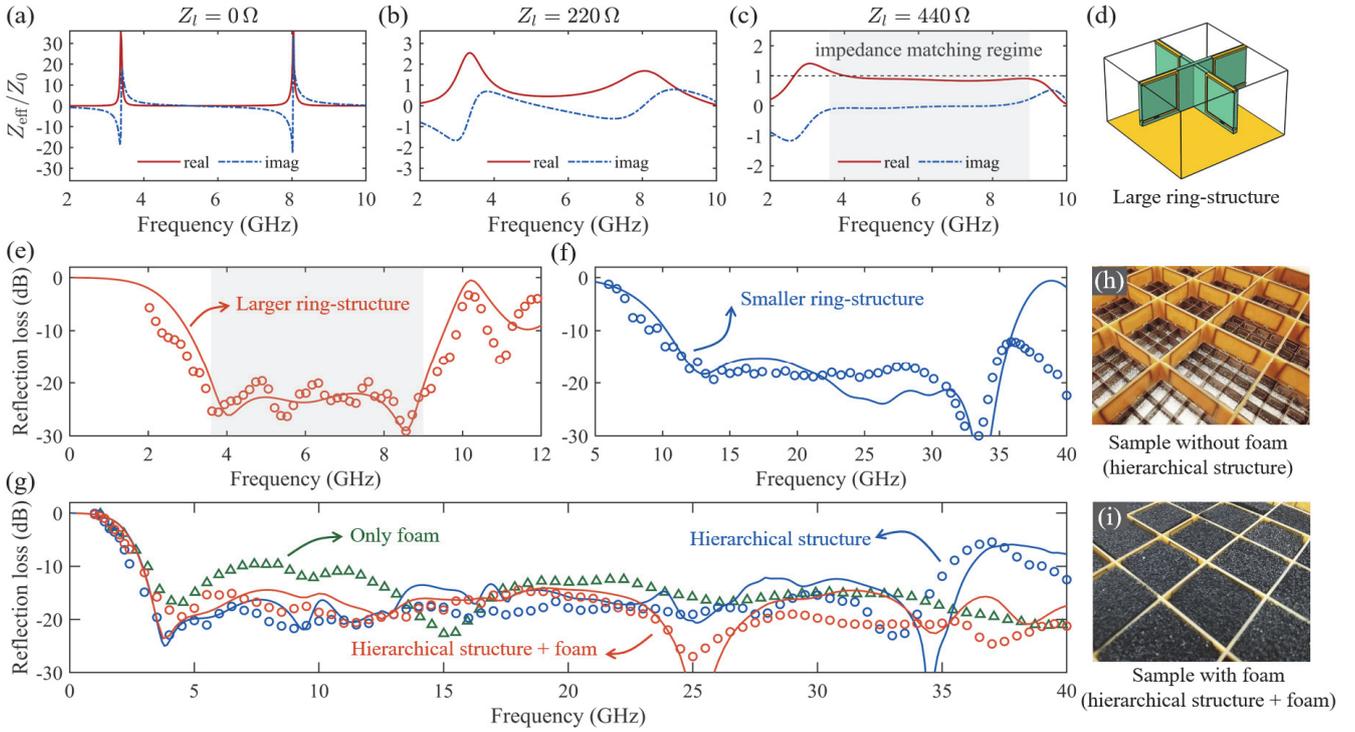

**Fig. 5 (a-c)** Results of dispersion engineering, by tuning the loaded resistance to be 0 Ω, 220 Ω and 440 Ω, respectively. **(d)** The unit cell of the large ring-structure. **(e)** The absorption performance with the optimal resistance of 440 Ω, achieved by a single large ring-structure. The solid lines are simulation results while the circles are experimental results. The grey shading indicates the impedance matching frequency regime. **(f)** The same for the small ring-structure. **(g)** The absorption of foam absorber (green), integrated hierarchical structure (blue) and hierarchical structure with foam (orange). **(h-i)** Image of the fabricated sample. Reused with the permission from Ref. [14].

discussed. Shen et al. [36] proposed concentric ring structures to create a triple-band absorber [see Fig. 4(b)]. Many similar works followed, with the aim of achieving perfect absorption by using multiple subwavelength resonant structures at discrete resonance frequencies [37-39]. Going a step further, Ding et al. [40] stacked a 20-layer metal structure to achieve larger mode density, thus fusing the discrete frequency bands into a continuous band [Fig. 4(c)]. Xiong et al. [41] also adopted multilayer metallic strips but used fewer layers to achieve similar broadband effects. The absorbers mentioned above were usually made of metallic structures, embedded in dielectric substrates. Some other materials were also explored to construct diverse microwave absorbers, such as liquid droplets [42-44], indium tin oxide (ITO) film [45, 46], graphene composites [47, 48], etc. Regardless of the specific composition of the materials, the ultimate goal is to support the resonant modes and to adjust the dissipation.

Sparse metallic structures generally have relatively strong resonant responses and are easy to fabricate with printed circuit board (PCB) technologies. However, owing to both the low resistance of the metals as well as the low ohmic loss of the substrate dielectrics, narrowband absorption features were difficult to avoid. This problem can be solved by introducing chip resistors to connect the metallic components. In principle, the resistance of the resistors can be designed to be any value, with infinity corresponding to the open-circuit, and infinitely small corresponding to the short-circuit. Therefore, dissipation in microwave systems is easy to regulate. By using chip resistors, a large class of works [49-51] achieved considerable success in broadband absorption, without many resonators as shown in Ref. [40]. Ye et al. designed a single-band perfectly-matched layer [52], and by adding resistively resonant units and optimizing their couplings, the same group extended the perfectly-matched layer effect to wider and also lower bands (from 1.1 to 2 GHz) [53] [see Fig. 4(d)].

Recently, a design recipe for an ultra-broadband microwave absorber was proposed by Qu et al. [14]. They discovered two magnetic resonances with high surface impedance [Fig. 5(a)] can be supported by an electrical dipole resonator in the form of a single metallic ring [Fig. 5(d)], placed close to a PEC boundary. By fine tuning the resistance loaded on the ring, the first-order and second-order magnetic resonances inevitably interact with each other [Fig. 5(b)], thereby leading to their coalescence into a broad impedance-matching band [Fig. 5(c)] (using the loaded resistors to manipulate the resonance dispersions is one manipulation approach of *dispersion engineering* [54]). Not surprisingly, the impedance matching condition leads to a -20 dB reflection loss between 3.6-9 GHz [Fig. 4(e)]. Furthermore, as the absorption frequency band is always inversely proportional to the relevant size of the absorber, hence if the size of the unit cell is scaled to 1/4 (keeping the optimal resistance unchanged), the absorption band of the smaller ring-structure will be around 4 times higher [see Fig. 4(f)]. Now, an opportunity to extend the absorption bands arises: the large and small ring-structures can be integrated along the wave incident direction to form a hierarchical structure [Fig. 4(h)], with its absorption spectra spliced as well [blue line and dots in Fig. 4. However, in such a case the diffraction inevitably appears due to the ultra-broadband coverage. To mitigate this problem, a conventional foam absorber was adopted to efficiently dissipate the additional diffraction beams [see the final combined structure in Fig. 4(i) and the reflection loss is plotted as orange line and dots in Fig. 4(g)]. Therefore, the final operating band (with a minimum of 90% absorption) was from 3 GHz to at least 40 GHz with a causality ratio $R_c = 1.05$ [14], which means that the causality optimality was achieved in this work. It is worth noting that the traditional material—sponge foam—also has quite good performance [green triangles in Fig. 4(g)]. This fact reveals that it is not easy for metamaterials to outperform sponge foams in practical applications, but the metamaterial absorbers possess the unique capability of frequency band tunability that is absent in traditional materials, which can be utilized to approach causal optimality in specified target frequency bands.

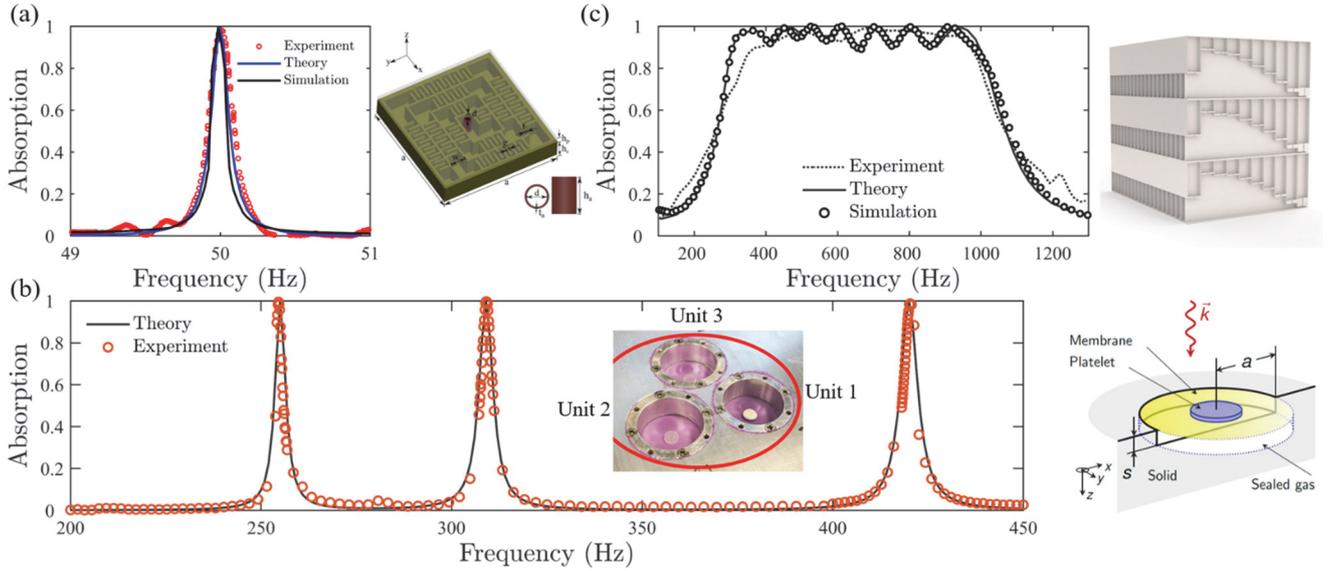

**Fig. 6** The performance of acoustic metamaterial absorbers and their structures. **(a)** A single-band low-frequency absorber with high absorption at 50 Hz. **(b)** A triple-band absorber with three integrated membrane resonators. The inset shows the fabricated integrated sample while on the right side, one unit cell for a single resonator is displayed. **(c)** A multilayer absorber with wider absorption band, enabled by stacked Helmholtz resonators. Reused with permissions from Ref. [76], Ref. [77] and Ref. [78].

In addition to moving toward wider band performance, microwave absorption metamaterials may also need to be lightweight [55], repairable [56], switchable [57], or optically transparent [45, 58] for practical applications. Also, how to achieve better oblique incidence performance is a problem worthy of further study. The concepts and principles used in microwave absorber can also be extended to higher bands such as terahertz [32, 33, 37, 59-61], infrared [33, 34, 62, 63] and visible light [64-68].

## B. Acoustic absorbers

Tuning the dissipation of the acoustic systems is not as flexible as that of microwave systems, since the viscosity coefficients of air and solid materials are relatively fixed. Therefore, the main method for broadband absorption in acoustic metamaterials is to increase the resonance mode density (route 2 in Fig. 3). However, in the case of acoustic waves we can always use hard boundaries to decouple different resonant modes; that is usually difficult in microwave absorbers with resonant elements owing to the non-negligible near-field couplings [17]; and we do not have the perfect magnetic boundary conductor (PMC) to confine the electromagnetic field as in the case of the acoustic waves by using the hard boundary.

Basic elements in airborne acoustic resonators can be Fabry-Pérot [20], Helmholtz [69], and membrane-type resonators [70], etc. By regulating the resonance frequency and oscillation strength, one can always achieve impedance-matching condition in principle by designing absorber's geometry. Early prototypes of single-band acoustic metamaterial absorbers date back to the micro-perforated panels (MPPs) investigated by Maa [71], who maximized the absorption bandwidth by optimizing the inner diameter and depth of the perforations. Recently, with more well-designed configurations, space-coiling [72-76] and membrane-type structures [77] have proven to support high absorption with thin thickness that is only one or even two orders magnitude of the relevant wavelength. Donda et al. [76] represented an acoustic metasurface targeting for 50 Hz sound, with only 13 mm thickness ($\sim\lambda/523$), as shown in

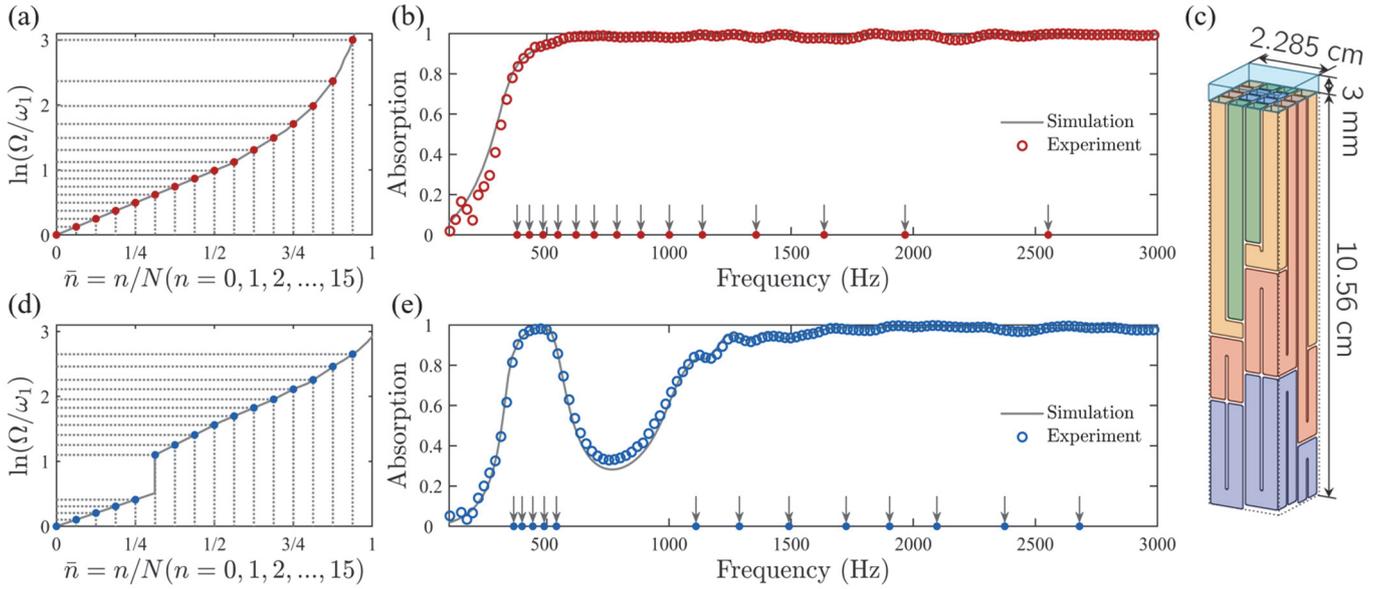

**Fig. 7 (a)** A scheme for arranging the Fabry–Pérot resonance frequencies (red dots), for broadband impedance matching. **(b)** The resulting absorption spectrum. The arrows indicate the distribution of the resonances in linear scale (only first-order resonances are plotted here). **(c)** A schematic of the sample, which was made of 16 air tubes, folded to be a compact structure. On the top, a 3mm-thick foam was placed to dissipate evanescent waves. **(d)** Another scheme for customized absorption bands. **(e)** The resulting customized absorption spectrum. The resonances and absorption were concentrated around 450 Hz and beyond 1100 Hz. Reused with permissions from Ref. [20].

Fig. 6(a). They took advantage of multi-coiled channels as an additional degree of freedom to achieve the low-frequency absorption. Ma et al. [77] utilized hybrid resonances to create a triple-band membrane-type absorber [see Fig. 6(b)]. They further demonstrated that the unusual large amplitude of membrane's vibration can be harnessed to effectively generate electrical energy with an efficiency of 23%. To further extend the absorption bandwidth, Jiménez et al. [78] constructed a stacked Helmholtz resonators (HRs) with different sized cavities [see Fig. 6(c)]. The dimensions of HRs were optimized by using the algorithm of sequential quadratic programming (SQP) to realize high absorption from almost 300 Hz to 1000 Hz. Many other works [79-83] also demonstrated both theoretically and experimentally that broadband behavior can be guaranteed as long as more resonant states are involved in building the absorber. Interestingly, under the action of natural selection, moths have also evolved micro-feathers of different sizes (as multiple resonators) on their wings to broadbandly absorb the ultrasonic detection signal from its natural enemy bats [84].

Yang et al. [20, 85] systematically developed a design scheme for the Fabry-Pérot resonances that can analytically predict the required resonance distribution for broadband impedance matching, shown in Fig. 7(a). The horizontal axis denotes the order of the resonators normalized by the total number $N$ while the vertical axis is the logarithm of first-order Fabry-Pérot resonance frequencies, normalized by the lowest $\omega_1 = 2\pi \times 343$ Hz. Grey line represents the ideal resonance distribution for continuum resonances ($N \to \infty$). But when it comes to the practical sample, only a finite number of resonators can be designed ($N = 16$) [see the red dots in Fig. 7(a) and the corresponding sample in Fig. 7(c)]. Since each Fabry-Pérot resonator can generate an infinite series of resonant states, absorption at high frequencies is usually easy to attain. The resulting performance is plotted in Fig. 7(b), with high absorption starting from 400 Hz. It was reported that the sample thickness (10.86 cm) is slightly larger than the causal minimal thickness (10.36 cm), thus $R_c = 1.048$. The same scheme can also be used to customize the absorption spectrum. As an example, by focusing on the absorption within the band of 350-550 Hz and 1200 Hz onwards, the resonance mode distribution can be updated by using the same algorithm for the targeted absorption bands. The latter is shown in Fig. 7(d). The resulting absorption and resonant modes are concentrated around 450 Hz and beyond 1100 Hz [see Fig. 7(e)]. The related sample thickness is 9.33 cm, also close to the causal optimality. This customizable strategy is especially useful for the noise source whose intensity distribution is mainly concentrated in certain bands. By using metamaterial absorbers, the space resources can be fully utilized to support the desired resonances. It is worth noting that recently more metamaterials absorbers have used alternative approaches, such as genetic algorithm [86], particle swarm optimization [87] and graph theory modelling [88] etc., to reach causal optimality using distinct structures.

We note that there is a class of acoustic absorbers [89-92] that operates in applications that require ventilation, i.e., where the rigid backing is replaced by impedance matching boundary. This is often the case in acoustic ductworks. By enhancing the oscillation strength of the resonances (stronger than that required by impedance matching), acoustic absorbers can be turned into sound insulating liners [24, 93-98] for targeted frequency band.

It should be noticed that the design concepts underlying the resonance-based metamaterials have already begun to diffuse to other systems, such as underwater acoustic absorbers [99-101] and elastic wave absorbers [102, 103], etc., holding broad applications potential.

# IV. METASTRUCTURES BEYOND THE CAUSAL LIMIT

The causality principle is perfectly general; it must hold regardless of its applications target. However, the derivation of Eq. (9) involves assumptions other than the causality principle, e.g., Neumann boundary condition ($\partial_x \phi |_{x=d} = 0$) on the backside of the sample, plus linearity of the sample response, passivity, and time-invariance etc. In addition, Eq. (9) is also noted to contain the static material properties $\mu_{eff}$ or $B_{eff}$. By either modifying the auxiliary assumption(s) or using materials with large static $\mu_{eff}$ or $1/B_{eff}$ values, it is entirely possible to respectively either mitigate or circumvent the stringent constraint imposed by the causal limit. In what follows, we describe some meta-structures that used different methods to achieve this goal.

## A. Mitigation approach

One can lower the factor $F$ in Eq. (9) so as to provide more absorption capacity with the same limited thickness resources, or to shrink the sample thickness without affecting the absorption spectrum. Usually, $F$ is slightly greater or equal to unity for most of the common EM and airborne acoustic metamaterials with high porosity (defined to be the reciprocal of $F$). In order to go beyond the original causal limit with $F \cong 1$, we should reduce the factor $F$ by introducing materials with large $\mu_{eff}$ or small $B_{eff}$.

In EM system, this can be done by using magnetic materials [104, 105], whose static permeability $\mu_{eff}$ is larger than that of air/vacuum $\mu_0$. However, since the large static magnetic permeability usually approaches one at microwave frequencies, simply using such materials can only make $R_c \gg 1$. Hence it is important that the large magnetic permeability values can persist to microwave frequencies. In the absence of such materials, one can use the large relaxational response at microwave frequencies for some magnetic materials to realize a smaller $F$ [106, 107] and thus good low-frequency absorption as well. We note that although the purely metallic structures can generate magnetic responses [108-110], at the static limit ($\omega \to 0$), the response is still non-magnetic.

For acoustic system, a small $F$ requires that the static bulk modulus $B_{eff}$ should be smaller than that of the incident medium $B_0$. This is feasible for underwater acoustic metamaterials. It was reported in Ref. [101] that an ultrathin but broadband underwater absorber was tailored to create a small $B_{eff}$ by using soft composite with high mass density. But whether the same idea can be used to design thinner absorbers in airborne acoustics remains an open question. It should be noted that the mitigation approach is still governed by the causal limit, i.e., Eq. (9) still holds but the tight constraint has been relaxed.

## B. Circumvention approach

Circumvention of the causal constraint means to violate some of the auxiliary assumptions used in the derivation of Eq. (9), thereby circumventing the causal constraint. There are several different methods under of this approach.

### 1. Coherent perfect absorption

Perhaps the best-known example is the coherent perfect absorbers (CPAs) [111]. In these devices, the amplitude, phase and frequency of the incident wave need to be accurately known a-priori to launch a coherent anti-symmetrical backward wave to interference with it and finally the energy of both forward and backward waves is totally dissipated within an ultrathin lossy layer. CPA was first discovered in the optical regime [112], and only later in the microwave regime [113]. The same concept was also simultaneously extended to acoustic systems [114-116].

### 2. Adding active components

Another way is to introduce active components that can achieve desired material properties with more degrees of freedom [117-119]. In particular, non-Foster elements [120-122] can be incorporated into the original passive EM structures to achieve better low-frequency absorption performance. In Ref. [120], measured results showed that the proposed microwave absorber achieved at least -10 dB reflection loss from 150 to 900 MHz with only 3 cm thickness. The adopted active non-Foster elements, based on negative impedance circuits (NICs), can create the response not available from the passive systems, which were also used to break the original performance limitation in not only wave absorption but also radiation [123, 124], cloaking [125], etc. But again, we note that the prerequisite information (i.e., amplitude) from the incident wave need to be known for setting the optimal power of the active components [120]. In addition, non-Foster elements should be carefully designed to avoid system instabilities. Although absorbers with active components have much better low-frequency performance, the price to pay is more sophisticated design and possibly higher production cost.

In recent years, with the deepening of the understanding of non-Hermitian systems [126], perfect absorption at the exceptional points has also attracted more attention due to the anomalous broadened absorption peaks [127], which has been confirmed by recent experiments [128, 129].

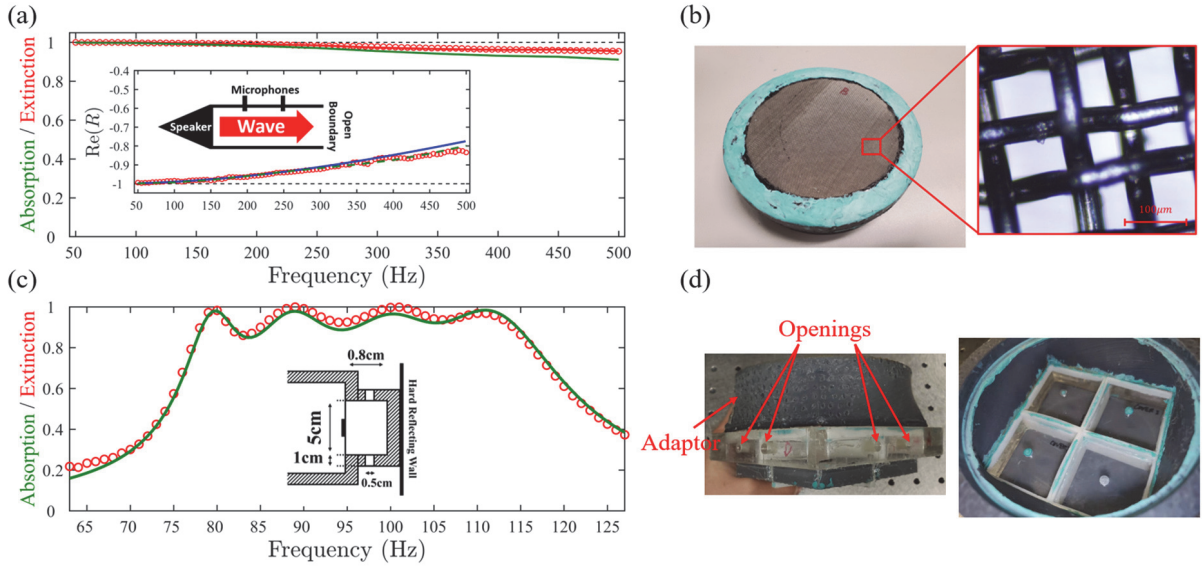

**Fig. 8** A meta-structure beyond the causal limit by altering the backside boundary condition, from hard to soft boundary. **(a)** Absorption and extinction spectra. The extinction is defined by $1 - |R|^2$, while the absorption $A = 1 - |R|^2 - |T|^2$. $R$ and $T$ are the reflection and transmission coefficients, respectively. The inset denotes the real part of reflection for an open tube without the sample. **(b)** A photo image of the metallic mesh used in **(a)**. **(c)** The absorption/extinction performance of the integrated membrane resonators. **(d)** A photo image of the fabricated sample with side openings to create as a partially soft boundary condition. Left and right photos denote side and top views. Reused with permissions from Ref. [134].

### *3. Temporal switching*

The third method is to violate the time-invariance assumption, since the causal limit is derived by Fourier transform to the frequency domain (implying the assumption of time-independent steady states). Hadad et al. [130] designed a spatiotemporally modulated device to realize non-reciprocity effects and found that the absorption can be much higher than the emission. Shlivinski et al. [131] proposed a temporal switching transmission-line between a source ($Z_S$) and load ($Z_L$). In a time-invariant system, the transmission efficiency is limited, especially when $Z_S \gg Z_L$ or $Z_S \ll Z_L$. However, it was found that if the incident medium's material properties can be switched to another state at the proper time, the efficiency can be even higher under the high-contrast case. In addition, the same group investigated the particular case in Ref. [132] when the load was replaced by perfect electric conductor (PEC) boundary, and the abrupt switching was improved by gradual switching. The results showed that the absorption can be larger than what the causal limit allows. Li et al. [133] developed a temporal switching absorber as well; its absorption spectrum can be customized according to practical need.

### *4. Altering the boundary condition*

The fourth method is to alter the boundary condition on the backside of the absorber from Neumann to Dirichlet boundary condition. The latter represents the perfect magnetic conductor (PMC) boundary condition for the EM case, and the soft boundary condition in the acoustic case.

By introducing acoustic soft-boundary-based absorber (ASBA), Mak et al. [134] detailed a new route for solving the age-old problem of low-frequency sound absorption. They used the end of an open tube to mimic the acoustic soft boundary, i.e., $p|_{x=d} \cong 0$, and placed stacked metallic meshes [Fig. 8(b)] in front of it to dissipate 99.99% of the incident energy at 50 Hz, and the sample thickness was optimized to be only 0.5 cm, which is only 7/10000 of the wavelength. The absorption was kept at a high level, beyond 90% from 50 Hz to 500 Hz [the green line in Fig. 8(a)]. It was shown that the same absorption using absorber backed by the hard boundary will need at least $d_{\min} \cong 80$ cm (an underestimated value since they only measured down to 50 Hz due to experiment limitation). In this experiment, the impedance at the open tube end will deviate from the perfect soft one with increasing of frequency due to the scattering caused by lateral modes [see inset in Fig. 8(a)]. But in the measured range (from 50 Hz to 500 Hz), the transmitted energy was less than 5%, which is negligible, especially when the frequency is low. Therefore, ASBA has un-paralleled performance in the low frequency limit, complementing the hard boundary condition that has advantages at high frequencies. The mechanism of ASBA is very similar to CPA, relying on a lossy layer with the proper thickness to achieve impedance-matching condition. However, as the soft boundary serves as an anti-symmetrical mirror for the incident wave field, it does not need the information from the incident wave.

In Ref. [134], it was further demonstrated that soft boundary can lower the impedance-matching frequency of the membrane-type absorber to below its resonance frequency, in contrast to the hard boundary case [77], whose impedance matching frequency should be higher than resonance frequency. To demonstrate this effect, four membrane resonators backed by a 2 cm-thick cavity with sidewall holes linked to open space were shown to display broadband low-frequency absorption (averaged value: 94%) from 80 Hz to 110 Hz [see Fig. 8(c)]. The sidewall holes were crucial for creating the boundary condition that can be characterized as "partially soft boundaries" [see Fig. 8(d)]. The causal limit in this case predicts a minimum thickness of at least 15.1 cm, much larger than the sample thickness.

## V. CONCLUDING REMARKS

In view of the rapid development of metamaterial absorbers, either under the causal framework or beyond, we conclude that at the laboratory stage, high absorption performance in most of the targeted frequency bands can now be realized. However, in the transition from scientific discoveries to large-scale applications, factors such as the manufacturing cost need to be considered. As many metamaterials are evolving towards evermore complex geometries [135], mass production cost is becoming an important consideration. We should note that 3D printing technology (mostly used in acoustic metamaterials) is still not a low-cost approach for mass production and it will face some limitations especially when multiple materials are involved. Also, when the geometric dimensions are small, accuracy can become a knotty problem. Developing a variety of low-cost manufacturing processes, customized for different types of metamaterials, may alleviate the cost problem of large-scale manufacturing in the future [136, 137], e.g., the printed circuit board (PCB) technology may be a promising cheap approach for fabricating EM metamaterials. Meanwhile, causal limit can also serve as a useful evaluation tool in applications. On the one hand, with the assistance of causal limit engineers can predict whether the given limited space is enough for targeted absorption performance if we use passive absorbers. Or, if the budget and practical situations allow, meta-structures beyond causal limit can be considered. On the other hand, causality ratio $R_c$ can be used to assess the performances of different absorbers, serving as one of the industry standards.

Metamaterial absorbers are envisioned to develop towards higher degree of performance customization in the future, with diverse structures that can fit into thin and compact volumes, crossing different wave systems. Machine learning and various optimization algorithms [138, 139] can assist in building the structure-performance networks for inverse design of the absorbers. The breakthrough in low-cost production technology will inevitably further accelerate the transition from laboratory samples to industrial products.

## ACKNOWLEDGEMENTS

P. S. acknowledges the support of A-HKUST601/18, Research Impact Fund R6015-18, and AoE/P-02/12 for this work.